\documentclass[aps,prl,reprint,groupedaddress, superscriptaddress]{revtex4-1}

\usepackage{graphicx}
\usepackage{amssymb}
\usepackage{amsmath}
\usepackage{bm}
\usepackage{hyperref}
\usepackage{lipsum}
\usepackage{bbold}
\usepackage{comment}

\begin{document}
\title{Commensuration torques and lubricity in double moire systems}
\author{Nicolas Leconte}
\affiliation{Department of Physics, University of Seoul, Seoul 02504, Korea}
\author{Youngju Park}
\affiliation{Department of Physics, University of Seoul, Seoul 02504, Korea}
\author{Jiaqi An}
\affiliation{Department of Physics, University of Seoul, Seoul 02504, Korea}
\author{Jeil Jung}
\email{jeiljung@uos.ac.kr}
\affiliation{Department of Physics, University of Seoul, Seoul 02504, Korea}
\affiliation{Department of Smart Cities, University of Seoul, Seoul 02504, Korea}

\begin{abstract}
We study the commensuration torques and layer sliding energetics of alternating twist trilayer graphene (t3G) and twisted bilayer graphene on hexagonal boron nitride (t2G/BN) that have two superposed moire interfaces.   
Lattice relaxations for typical graphene twist angles of $\sim 1^{\circ}$ in t3G or t2G/BN are found to break the out-of-plane layer mirror symmetry, give rise to layer rotation energy local minima dips of the order of $\sim 10^{-1}$~meV/atom at double moire alignment angles, and have sliding energy landscape minima between top-bottom layers of comparable magnitude.
Moire superlubricity is restored for twist angles as small as $\sim 0.03^\circ$ away from alignment resulting in suppression of sliding energies by several orders of magnitude of typically $\sim 10^{-4}$~meV/atom, hence indicating the precedence of rotation over sliding in the double moire commensuration process.
\end{abstract}

\maketitle

\textbf{Introduction} -- Experimental realization of moire materials by layering two graphene-like 2D material systems with different lattice constants and/or with a finite twist angle~\cite{Dean:bv, Geim2013} have been furthered by combining two moire patterns to form the so-called double or super moire systems~\cite{Finney2019, Wang2019nl, Anelkovi2020, PhysRevB.81.125427, Wang2019bis, PhysRevLett.127.166802, Leconte2020, https://doi.org/10.48550/arxiv.2209.12154,Lee2020}. 
Crystals with long moire pattern periods allows to access certain physical observables 
at low magnetic fields and gate carrier densities that would not be accessible otherwise~\cite{Hunt2013,Dean2013,Ponomarenko2013}. 
This behavior can be used for example in $30^\circ$ twist-angle quasi-crystals~\cite{Yao2018, Deng2020, PhysRevB.99.165430, Yu2019, PhysRevB.100.081405} to generate moire quasicrystals at experimentally accessible charge carrier densities using h-BN encapsulated graphene~\cite{Leconte2020, Crosse2021}. 
Most theories describing double moire systems have so far used the simplest commensurate double moire geometries to explain the observed physics.
For instance, in alternating twist trilayer graphene (t3G) with twisted middle layer giving rise to two aligned moire patterns, important in the context of flat band superconductivity
~\cite{cao2018,Park2021,Hao2021,li2019, Cea2019, PhysRevResearch.2.022010, 1901.09356,  Liu2020, 2203.09188, jeongminpark2021, Hao2021, jeongminpark2022},
the most stable geometry corresponds to superposed top and bottom layers.
Likewise commensurate double moire geometries have been assumed when studying twisted bilayer graphene on h-BN where a spontaneous anomalous Hall effect was measured~\cite{PhysRevB.103.075122, Long2022, PhysRevB.102.155136, doi:10.1126/science.abd3190, doi:10.1126/science.aay5533, doi:10.1126/science.aaw3780}.
However, the assumption that equal period and aligned double moire pattern systems are energetically favored has not been yet confirmed. 

Here we analyze the atomic structure of t3G and twisted bilayer graphene on hexagonal boron nitride (t2G/BN) 
to show that double moire systems generate torques that tend to lock the systems into commensurate moire patterns 
and favor a specific sliding geometry.
We find that mirror symmetry breaking layer corrugations are required for a correct total energy minimization, including the $\overline{\rm AAA}$-stacked t3G systems where we use the overline to indicate relative sliding geometries between the layers regardless of their twist angles.
We observe that the specific sliding atomic structures between the top and next nearest layer only matters when the moire patterns are commensurate, since the energies are almost sliding-independent for incommensurate moire patterns leading to superlubricity away from commensuration, 
similar to the superlubricity behavior in single interface twisted bilayer systems studied in the literature~\cite{doi:10.1021/acsanm.1c01540,Bai2022,PhysRevLett.92.126101,Wang2019, doi:10.1021/nl204547v,PhysRevLett.100.046102,1908.04666}.

\begin{figure*}[tbhp]
\begin{center}
\includegraphics[width=0.85\textwidth]{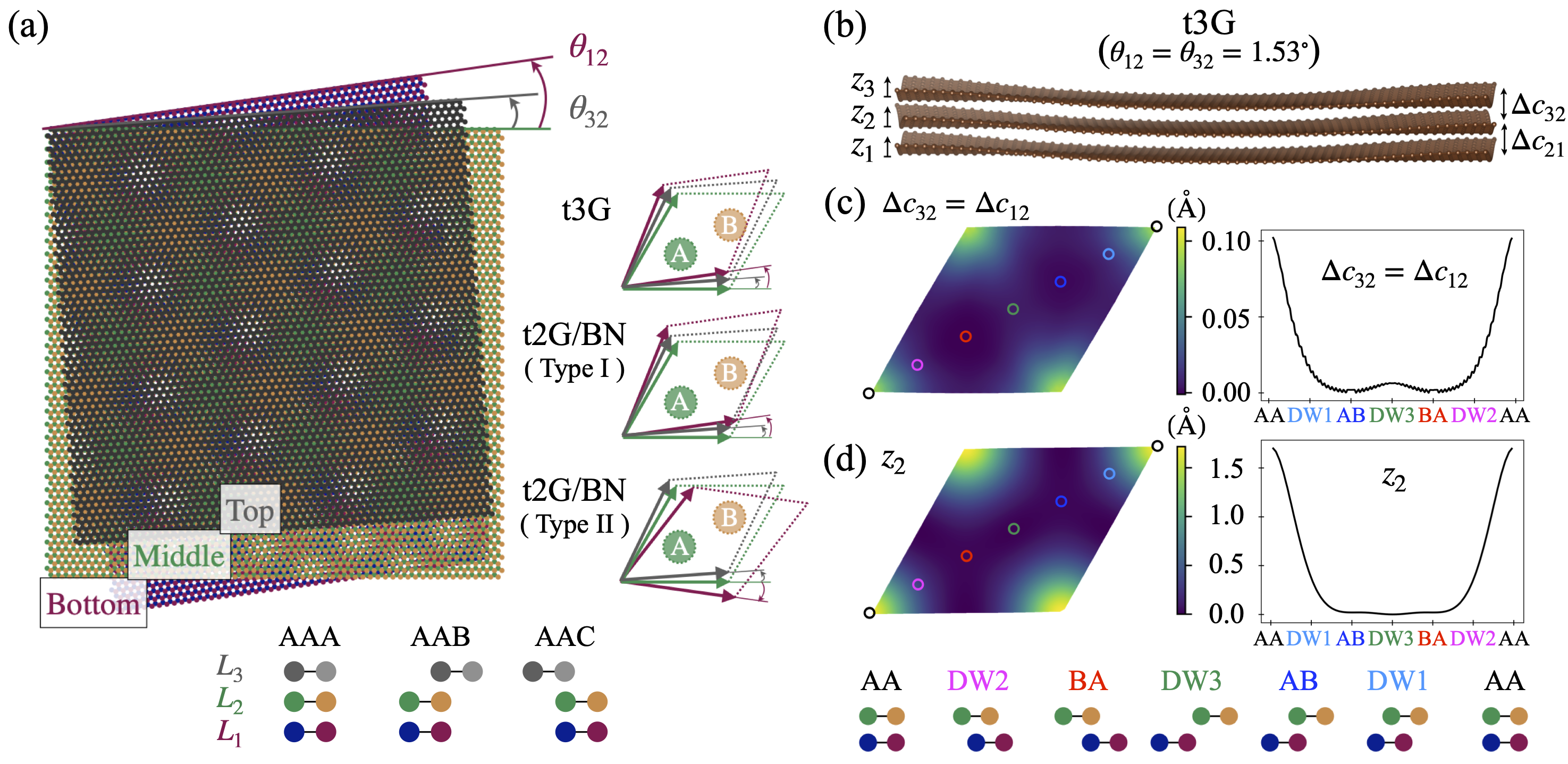}
\caption{(color online) (a)
Schematic figure illustrating the different t3G and t2G/BN systems considered here. 
$\theta_{ij} = \theta_i - \theta_j$ are the twist angles of the $i$-th layer with respect to the $j$-th layer where their particular choices can lead to commensurate and incommensurate moire patterns.
(b) Corrugation effect that breaks the mirror-symmetry of $\overline{\rm AAA}$ stacked t3G with $\theta_{12} = \theta_{32} = 1.53^\circ$.
(c) Interlayer distance differences at different local stacking positions and (c) bending corrugations in the middle layer 2. 
The sketches at the bottom of the panels indicate the specific local stackings AAA, AAB and AAC for the three-layer systems while the two-letter conventions refer to local sliding between two layers. Gray, green and orange colors refer to the A and B sublattices of graphene.
Blue and purple refer to the boron and nitrogen atoms in hBN in t2G/BN or the A and B sublattices in a t3G system. 
}
\label{fig:schematic}
\end{center}
\end{figure*}

\textbf{Systems and methods} -- We consider double moire systems with two moire interfaces consisting of graphene trilayers and t2G on hexagonal boron nitride.  
We illustrate schematically in Fig.~\ref{fig:schematic} (a) all three different systems considered, namely t3G, t2G/BN type I and type II depending on the twist angle sense of the bottom hBN layer contacting graphene.
The layer numerals $1$, $2$ and $3$ correspond to bottom, middle and top layers respectively. 
The middle layer 2 is taken as the reference frame with zero twist angle and we use $\theta_{12}$ and $\theta_{32}$ labels to represent the actual twist angles of the bottom and top layers.
The atomic structure relaxation is carried out using LAMMPS~\cite{Plimpton1995}. 
We use the REBO2 force-field~\cite{Brenner_2002} for the intralayer interactions of graphene and EXTEP~\cite{PhysRevB.96.184108} for those of hexagonal boron nitride, whose equilibrium geometries 
are $a_{\rm G} =$2.4602\AA~ and $a_{\rm BN} =$2.50576\AA~ respectively. 
The interlayer force fields are based on EXX-RPA-informed~\cite{PhysRevB.96.195431} DRIP~\cite{PhysRevB.98.235404} parametrizations~\footnote{The CBN\_RPA.drip and CBN\_LDA.drip potential files to be used with LAMMPS are available at \url{https://github.com/gjung-group/real-space_relaxation_electronic-structure-calculations} and these can be used with the input file from the DRIP example folder in the LAMMPS directory.} and we used both the fire and CG minimization scheme~\cite{POLYAK196994}
with a time step of $0.0001$~ps for the former and $0.001$~ps for the latter and a stopping tolerance on the forces of $10^{-5}$~eV/\AA.
In order to capture the tiny angle variations with respect to the doubly commensurate angles 
we use the approach outlined in Ref.~\cite{PhysRevB.106.115410} and Ref.~\cite{Hermann:2012dy} to find the commensurate cell for a double moire system, see Appendix A for details and summary of the integers representing our commensurate cells. 
The incommensurate moire systems are approximated by taking commensurate simulation cells containing several repetitions the moire unit cells or moirons. Due to the small angle differences and the high sensitivity of the results on internal strains, we are at times bound to choose large simulation systems containing millions of atoms. The lattice constant variation tolerance is capped at $0.03\%$ to minimize internal strains that lead to energy differences of the order of $0.005$~meV/atom that are between one to two orders of magnitude smaller than the energy differences of the order 0.1~meV/atom required to resolve the local minima in our energy curves.
The stability of the atomic structure relies on the total energy given as the sum 
\begin{eqnarray}
E_{\rm tot} = E_{\rm el} + E_{\rm pot}
\label{totalEnergy}
\end{eqnarray}
where we can distinguish the elastic energy $E_{\rm el} =  \sum_{i} E_{\rm el}^{i} /2$ that resists the deformation due to the strains, 
and the potential energy $E_{\rm pot} = \sum_{i} E_{\rm pot}^{i} / 2$ that triggers the formation of the moire pattern strains taken as sum of contributions from each atomic site $i$ and where the division by 2 accounts for double counting.
We can define the local elastic $E_{\rm el}^i$, potential $E_{\rm pot}^i$ and interface $E_{\rm IF_{mn}}^i$ energies as 
\begin{eqnarray}
E_{\rm el}^i &=& \sum_{j\in \text{layer i}} \phi^{ij}    \label{elastic} \\
E_{\rm pot}^i &=&  \sum_{j\notin \text{layer i}} \phi^{ij}  =  \sum_{j\in \text{ any layer}} \phi^{ij} - E_{\rm el}^i \label{potential} \\
E_{\rm IF_{mn}}^i &=& \sum_{
    \scriptsize
    \begin{tabular}{c}
      $j \notin \text{layer i}$ \\
      $j\in \text{layer n or m}$ 
    \end{tabular}
    } 
\phi^{ij}_\text{mn}    
\label{potElIF}
\end{eqnarray}
where $\phi^{ij}$ represents the pair-wise potentials between atoms $i$ and $j$. 
Our calculations show that the elastic energy contributions in Eq.~(\ref{elastic}) are about one order magnitude smaller than the potential and interface energies in Eqs.~(\ref{potential}-\ref{potElIF}), 
and therefore is only a small fraction of the total energy in Eq.~(\ref{totalEnergy}) 
that dictates the stability of our systems.  
The interface energies are essentially the potential energies referred to a particular pair of layers.
It will be interesting to note how this interface energy changes from one system to another by examining the interface energy differences for two different relaxed atomic structures
\begin{equation}
\Delta E_{\text{IF}_{mn}}({\bm r}) = E^{3L}_{\text{IF}_{mn}}({\bm r}) - E^{2L}_{\text{IF}_{mn}}({\bm r})
\label{deltaIF}
\end{equation}
where $E^{3L}_{\text{IF}_{mn}}$ at a given point is obtained relaxing simultaneously all three 
layers of t3G and then considering the bilayer atomic positions for the considered $mn$ interface, 
while the $E^{2L}_{\text{IF}_{mn}}$ interface energy is obtained using the t2G relaxed atomic positions of the two $mn$ layers that form the interface.
In Eq.~(\ref{deltaIF}) we have removed the $i$-index dependence in $E^{i}_{\text{IF}_{mn}}$ in Eq.~(\ref{potElIF}) using the position vector ${\bm r}$ instead by interpolating the data 
from the closest $i$-sublattice points.

Another quantity of interest is the torque constant that we define as the derivative of the total energy as a function of twist angle similar to the proposals in Refs.~\cite{PhysRevB.41.11837, doi:10.1021/nl204547v,1908.04666} but focusing here on the rotation of the top layer with respect to the middle layer
\begin{equation} 
    k_{\pm} = 
    \frac{d E_{\rm tot}}{d \theta_{32}}
    \label{torqueConst}
\end{equation}
where its positive or negative values tend to either reduce or increase the value of $\theta_{32}$ towards the commensurate moire geometry. 

\begin{figure*}[tbhp]
\begin{center}
\includegraphics[width=0.9\textwidth]{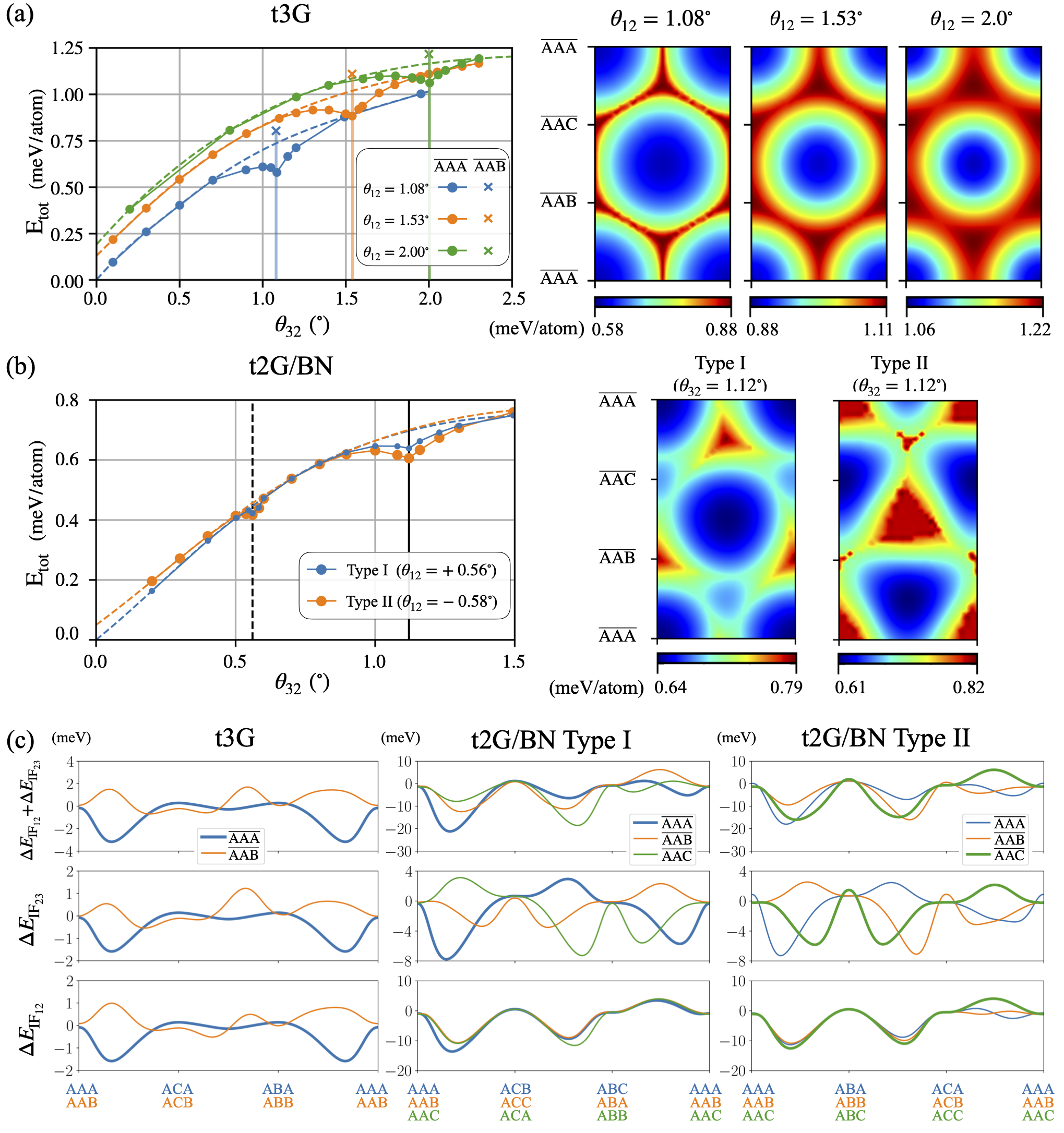}
\caption{(color online) (a) 
Energies of t3G double moire systems for three different values of $\theta_{12}$ where $\theta_{32}$ is varied from $\sim0.1^\circ$ to $\sim2.2^\circ$. For t3G systems the commensurate moire patterns are obtained when $\theta_{12}=\theta_{32}$ (indicated by vertical lines) and leads to a local energy dip. The x-symbols represent the total energies when the top interface has $AB$-stacking. The dashed lines are obtained by using polynomial interpolation away from the commensuration dips.  
On the right hand we show total energies for different sliding of the top layer for select commensurate angles $\theta_{12}=\theta_{32}$. (b) Similar plots for t2G/BN (Type I and Type II) where the total energy dips are obtained for two different values of $\theta_{32}$, corresponding to angles where $L_{12} = L_{32}$ (solid vertical line) and $L_{12} = L_{32} /2$ (dashed vertical line). 
(c) Intepolated interface energy differences $\Delta E_{\rm IF_{mn}}({\bm r})$ of Eq.~(\ref{deltaIF}) 
for the 3 systems considered illustrating the local energy gain/penalty when a single moire comes into contact with a second moire interface. The energy difference densities are plotted along a straight line that connects the opposite diagonal corners of the moire cell through different local stacking positions where the most stable stacking arrangement is highlighted with a thicker line. The lower the energy, the more stable is the double moire geometry locally. 
The negative energies correspond to a stabilizing gain in energy while positive energies indicate a destabilizing energy penalty. The interface energy differences, and therefore the potential energy differences, make up the dominant contributions of the total energy differences since the elastic energy contributions are one order of magnitude smaller.}
\label{fig:t3G}
\end{center}
\end{figure*}

\textbf{Results} -- 
We begin by testing the main assumption of our study for t3G, namely that angles leading to commensurate moire systems will be energetically more favorable than angle combinations leading to incommensurate moire systems. 
For this purpose we fix the angle $\theta_{12}$ between the two bottom layers to three select values of $\theta_{12} = 1.08^\circ$, $1.54^\circ$ and $2.0^\circ$ to sample typical largest magic angles values in t$N$G systems for $N = 2, 3, \infty$~\cite{Leconte_2022} while we vary $\theta_{32}$ between the two top layers from $0.1^\circ$ to $2.3^\circ$. Commensuration for alternating twist is naturally achieved when $\theta_{12} = \theta_{32}$ since the lattice constants of all three layers are equal. 
The find numerically that the relaxed atomic structures generally favor $z$-axis corrugations, 
where Fig.~\ref{fig:schematic}(c) shows the typical local stacking-dependent interlayer distance 
variation of the order of $\sim 0.1$~\AA\ while Fig.~\ref{fig:schematic}(d) illustrates 
for the middle layer the $z$-axis corrugations of the order of $\sim 1.75$~\AA\, roughly one half of the average interlayer distance.

The main results of this work are shown in Fig.~\ref{fig:t3G}(a) where we plot the total energy per atom as a function of $\theta_{32}$ after the atomic relaxation. 
For all three $\theta_{12}$ values considered, we find total energy local minima at the commensurate twist angles provided that we allow a bending corrugation as illustrated in Fig.~\ref{fig:schematic}(b). In fact, the total energies reported in Table I indicate that the mirror-symmetric t3G with a completely flat middle layer has a higher total energy compared to the corrugated atomic structure.
There we list the integrated total, elastic, potential, and interface energies taken as the sum over all atomic sites $i$, with proper double counting correction, of the energies given in Eqs.~(\ref{totalEnergy}-\ref{potElIF}).
In Fig.~\ref{fig:t3G}(a) we show that the commensurate moire systems are most stable in the $\overline{\rm AAA}$-stacking when there is no relative sliding between the three layers.
In fact, the energy difference between the $\overline{\rm AAA}$ at the local minima dip and $\overline{\rm AAB}$ maxima represented is equal to $0.18$~meV/atom near $1^{\circ}$ and drops to $0.06$~meV/atom for an angle of $3.47^\circ$ (not shown here) and are comparable to the magnitude of the energy dips due to the rotation. Our sliding dependent total energy plots indicate that there are barrier-free sliding paths leading to the global minimum at $\overline{\rm AAA}$-stacking.

We then show a similar analysis for t2G/BN where we fix the substrate angle between G and hBN at $\theta_{12} = 0.56^{\circ}$ for type I or $-0.58^\circ$ for type II and we allow $\theta_{32}$ between both graphene layers to change up to a value of 1.5$^{\circ}$ to achieve moire periods that satisfy $p L_{tBG}^M = q \, L_{tGBN}^M$~\cite{PhysRevB.103.075423} 
where $p, \, q$ are integers.
For $q=1$ we observe two dips in the energy curve that correspond to $\theta_{32}=0.56^\circ$ for $p=2$ with a G/G moire pattern period twice as large as the G/BN and $\theta_{32}=1.12^\circ$ for $p=1$ with equal periods.

The details of the atomic relaxation giving rise to total energy dips in Fig.~\ref{fig:t3G}(a) and (b) for t3G and t2G/BN can be further understood through the line plots in Fig.~\ref{fig:t3G}(c)
calculated using Eq.~(\ref{deltaIF}) where we focus on the interface energy difference between 3-layer and 2-layer relaxed systems.
We can thus quantify the energy gain/penalty (negative and positive values, respectively) one achieves when putting two moire patterns in contact with each other. 
For t3G the most stable 3-layer configuration is the $\overline{\rm AAA}$-stacking  that we achieve when the unstable AA-stacking of each bilayer are stacked on top of each other and the stable AB/BA stackings are simultaneously stacked on top of each other. 
The interface energy plots in Fig.~\ref{fig:t3G}(c) illustrate the energy differences between single and double moire pattern atomic structures. 
For $\overline{\rm AAA}$-stacking the interface energy difference plot shows that its energy gain is most pronounced at local stacking regions in between the AAA and ACA/ABA regions. 
In contrast the $\overline{\rm AAB}$-stacking has a generally unfavorable interface energy for most local stackings. 

The double moire energetics becomes more complicated when we have an heterogeneous interface in t2G/BN
where the most stable (unstable) AB/BA (AA) local stacking geometries in t2G
combines with the AC (AA/AB) stackings of GBN.
For type I trilayers the energetically stable geometry is $\overline{\rm AAA}$ while 
for type II trilayers it is $\overline{\rm AAC}$. 
We notice that the strongest local energy penalties due to interfering moire patterns do not necessarily happen at the high symmetry local stacking geometries and the global energy minimization does not follow simple rules of thumb consisting in combining together the local stackings that are energetically most favorable or unfavorable as in the t3G case. 

\begin{figure}[tbhp]
\begin{center}
\includegraphics[width=0.9\columnwidth]{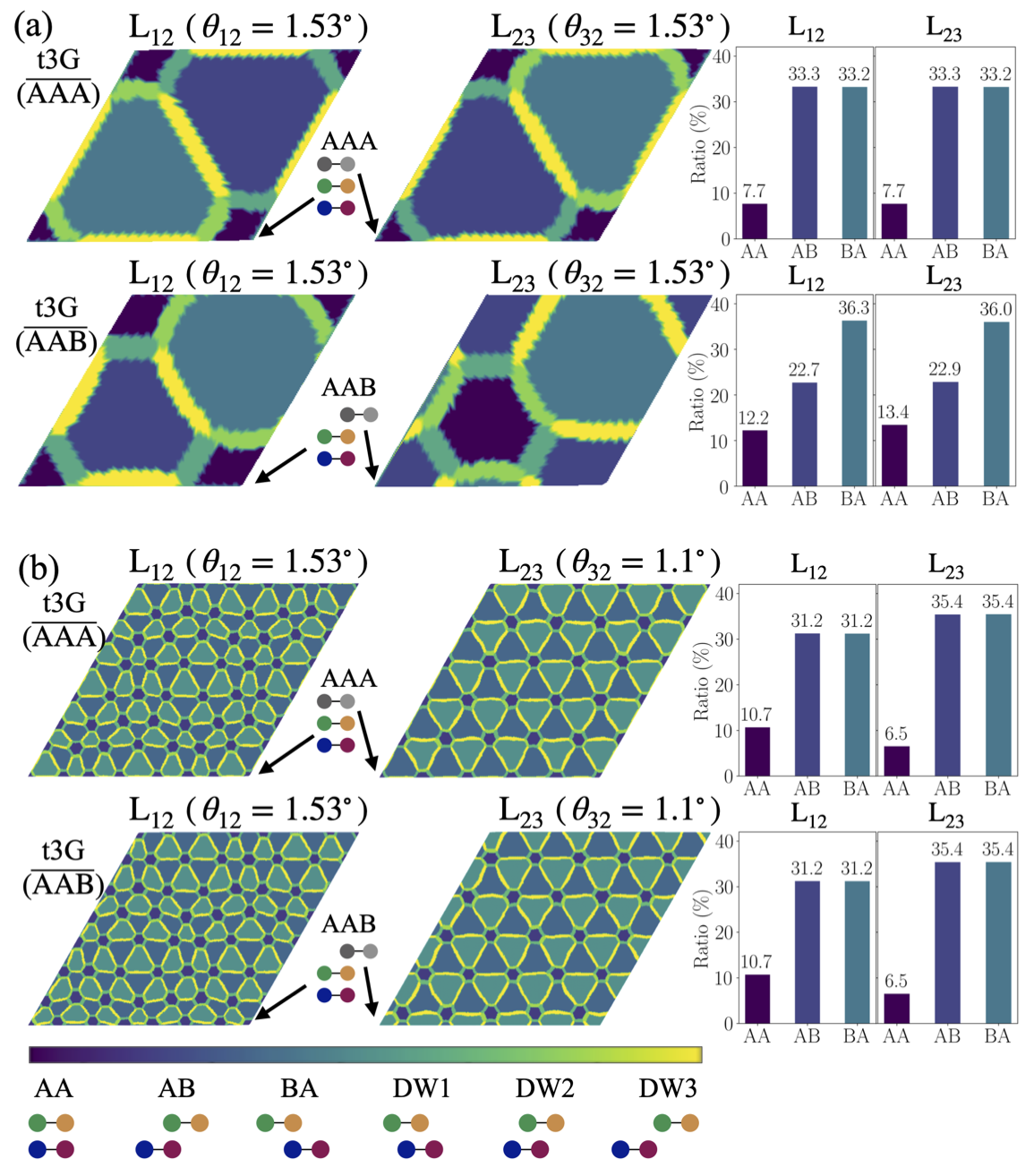}
\caption{ Illustration of the stacking redistribution in t3G for two double moire configurations obtained using the conventions outlined in Ref.~\cite{1901.09356} using the displacement vectors between layers $m$ and $n$. 
(a) Local stacking maps for $\overline{\rm AAA}$ and $\overline{\rm AAB}$ stacking geometries for $\theta_{12} = \theta_{32} = 1.53^{\circ}$, and the ratios of the local stacking areas of the left panels. The $\overline{\rm AAA}$ stacking shows a smaller AA local stacking area due to greater local rotations that reduces this energetically unfavorable stacking when compared with the $\overline{\rm AAB}$ stacking going from 7.7\%  to 12.2\% respectively.
(b) Local stacking maps for $\theta_{12} = 1.53$ and $\theta_{32} = 1.1^\circ$ for sliding geometries and corresponding local stacking area ratios. 
Although the relaxation profiles are different, the ratios for local AA, AB and BA stackings remain nearly the same for different sliding geometries. }
\label{fig:stacking}
\end{center}
\end{figure}

\begin{table}[tb]
\footnotesize
\begin{tabular}{|l|c|c|c|c|}
 \hline
                   & E$_\text{tot}$  & E$_\text{el}$ & E$_\text{IF12}$  & E$_\text{IF23}$   \\
                   \hline
t3G ($\overline{\rm AAA}$)          &    $-$7.42537          &    $-$7.39458      &          $-$0.01542     &     $-$0.01541                      \\ \hline
t3G ($\overline{\rm AAA}$, ms)          & $-$7.42534              &  $-$7.39459             &       $-$0.01538         & $-$0.01538                               \\ \hline
t3G ($\overline{\rm AAB}$)           &     $-$7.42510          &       $-$7.39471        &    $-$0.01521            &       $-$0.01516                        \\ \hline
t2G/BN-I ($\overline{\rm AAA}$)  & $-$7.19472 &  $-$7.16454 &  $-$0.01458   & $-$0.01559     \\\hline
t2G/BN-I ($\overline{\rm AAB}$)  & $-$7.19451 &  $-$7.16466 & $-$0.01442    &  $-$0.01542   \\\hline
t2G/BN-I ($\overline{\rm AAC}$)  & $-$7.19467 & $-$7.16455  & $-$0.01456    &  $-$0.01556    \\\hline
t2G/BN-II ($\overline{\rm AAA}$) & $-$7.19561 & $-$7.16554  & $-$0.01452   &   $-$0.01556  \\\hline
t2G/BN-II ($\overline{\rm AAB}$) & $-$7.19562  & $-$7.16553  &  $-$0.01453   &  $-$0.01556    \\\hline
t2G/BN-II ($\overline{\rm AAC}$) & $-$7.19581 &  $-$7.16556 &  $-$0.01463   &  $-$0.01563   \\\hline

\end{tabular}
\caption{
Sliding dependent total,  elastic, and interface energies in for $\theta_{12}=\theta_{32}=1.5385^\circ$ for t3G,
$\theta_{12}=0.56^{\circ}, \theta_{32}= 1.12^{\circ}$ for t2G/BN-I, and $\theta_{12}=-0.58^{\circ}, \theta_{32}= 1.12^{\circ}$ for t2G/BN-II for systems containing 8322, 15490 and 15492 atoms respectively. These numbers are used to renormalize and report the energies in eV/atom. We note that the mirror-symmetric (ms) geometry has a higher total energy by 0.03~meV/atom. 
}
\end{table}
The conclusions on stability energetics that we can draw based on the energy density difference line cuts in Fig.~\ref{fig:t3G} are consistent with the interface and total energies that predict the $\overline{\rm AAA}$, $\overline{\rm AAA}$ and $\overline{\rm AAC}$ stacking as the most stable geometries for t3G, t2G/BN type I and t2G/BN type II, systems respectively.
To further explore the stabilization of the commensurate moire systems versus the incommensurate ones we show in Fig.~\ref{fig:stacking} the local stacking distribution maps for $\overline{\rm AAA}$ and $\overline{\rm AAB}$ stackings of t3G following the conventions outlined in Ref.~\cite{PhysRevB.106.115410} for each one of the interfaces.
Here we can compare how the local stacking distributions change for the commensurate and incommensurate moire interfaces, and how the favorable and unfavorable sliding geometry influences the strain profiles. 
For commensurate moire patterns and stable $\overline{\rm AAA}$-stacking the energetically unfavorable AA local stacking that makes about 7.7\% of the total area increases to 12.2\% for the less stable $\overline{\rm AAB}$-stacking.
As soon as we move away from the same period commensurate moire geometries and have different moire pattern periods,
for instance the combination of $1.54^\circ-1.1^\circ$ with 1/7 moire length ratios
or $1.54^\circ-1.5^\circ$ with 1/39 ratios that can be considered approximations of the incommensurate moire patterns. 
For these geometries the relative distribution of the AA, AB or BA local stacking areas become practically insensitive to the relative sliding of the top layer resulting in almost the same local stacking area ratios for both $\overline{\rm AAA}$ and $\overline{\rm AAB}$ stackings. We can thus conclude that the sliding-dependent energy landscape of the outer layers will have non-negligible gradients only when we have equal period commensurate moire patterns.
Sliding energy landscapes of $10^{-1}$~meV/atom for commensurate double moire systems where $\Delta \theta = \theta_{32} - \theta_{12} = 0^\circ$ 
quickly drops to $10^{-4}$~meV/atom even for marginally small twist angles away from commensuration  
of $\Delta \theta \simeq 0.03^\circ$ to down to $10^{-5}$~meV/atom for $\Delta \theta \simeq 0.5^\circ$,
indicating high moire superlubricity away from exact commensuration.
This is consistent with the fact that both structures shown in Fig.~\ref{fig:stacking} for the $1.54^\circ-1.1^\circ$ combination have nearly the same local stacking area ratios, indicating in turn that they will have weak interlayer sliding force gradients. 

The torque constants obtained using Eq.~\ref{torqueConst} from the total energy curves are summarized in Table II together with the binding energies $E_b$ defined as the energy gain achieved due to commensuration that we define as the difference between dashed interpolated line and the actual total energy at commensuration in Fig.~\ref{fig:t3G}(a) and (b). 
The calculated torque magnitudes are generally larger for smaller angles and the signs of $k_+$ and $k_-$ torques tend to bring the incommensurate moire systems back to commensuration as a positive/negative torque value will decrease/increase the $\theta_{32}$.
\begin{table}[]
\small
\footnotesize
\label{torque}
\begin{tabular}{|c|c|c|c|}
 \hline
  \multicolumn{4}{|c|}{ t3G ($\theta_{12}=\theta_{32}$) }\\ \hline
 $\theta_{32}\ (^{\circ})$ &   $1.08$     &  $1.53$     & $2.00$  \\ \hline
\begin{tabular}[c]{@{}c@{}} $ k_{-}(\theta_{32}) $ \\ (meV/(atom$\cdot{\rm rad}$)) \end{tabular}  & $-$42.12   &  $-$18.34  &  $-$8.513     \\ \hline
\begin{tabular}[c]{@{}c@{}} $ k_{+}(\theta_{32}) $ \\ (meV/(atom$\cdot{\rm rad}$)) \end{tabular}   &  75.06     &   53.04   &  36.85     \\ \hline
\begin{tabular}[c]{@{}c@{}} $E_b(\theta_{32})$ \\ (meV/atom)\end{tabular}    & 0.156322    &  0.135053  &  0.103796  \\  \hline
 \end{tabular}
 \begin{tabular}{|c|c|c|c|c|}
  \hline
t2G/BN & \multicolumn{2}{|c|}{ \begin{tabular}[c]{@{}c@{}} Type I \\ ($\theta_{12}=+0.56^{\circ}$)\end{tabular} }
&  \multicolumn{2}{|c|}{ \begin{tabular}[c]{@{}c@{}} Type II \\ ($\theta_{12}=-0.58^{\circ}$)\end{tabular} }\\ \hline
$\theta_{32}\ (^{\circ})$ &   $0.56$     &  $1.12$   &   $0.56$     &  $1.12$      \\ \hline
\begin{tabular}[c]{@{}c@{}} $ k_{-}(\theta_{32}) $ \\ (meV/(atom$\cdot{\rm rad}$)) \end{tabular}     & $-$40.01   & $-$9.69     & $-$7.124  & $-$14.33      \\  \hline
\begin{tabular}[c]{@{}c@{}} $ k_{+}(\theta_{32}) $ \\ (meV/(atom$\cdot{\rm rad}$)) \end{tabular}    &    48.01   &   35.52    &    54.54 &    40.46      \\  \hline
\begin{tabular}[c]{@{}c@{}} $E_b(\theta_{32})$ \\ (meV/atom)\end{tabular}    & 0.024273   & 0.058352    & 0.039000  & 0.094661      \\ \hline
\end{tabular}
\caption{Torque constants $k_{\pm} = dE_{\rm tot} / d \theta_{32}$ in units of meV/(atom $\cdot$ rad) as defined in Eq.~(\ref{torqueConst}) evaluated to the left ($k_-$) and right ($k_+$) of the respective local minima at the commensurate angles $\theta_{32}$, and the binding energy $E_{b}(\theta_{32})$ estimated as the difference between the smoothly interpolated polynomial curve and the respective minima in Fig.~\ref{fig:t3G}(a)-(b).}
\end{table}

The analysis we have presented so far relied on free-standing trilayer systems. In order to assess the impact a substrate would have on our results, we have performed the following checks. For t3G, we have checked that adding a rigid hBN substrate layer with a twist angle of $3.41^\circ$ dampens but does not completely remove the bending corrugation observed for the mirror symmetry broken commensurate case. Indeed, the maximum bending corrugation goes down for the suspended t3G system from $1.70$~\AA\ as seen in Fig.~\ref{fig:schematic}(d) to $0.25$, $0.35$, and $0.45$~\AA\ for $L_1$, $L_2$ and $L_3$, respectively, when we add a rigid substrate layer in contact with $L_1$. If we don't fix the additional substrate layer, the bending corrugation actually increases by up to $3.79$~\AA, hence a realistic substrate simulation, outside of the scope of this study, involving many more layers~\cite{Leconte_2022, Mandelli2019-au} would probably give a maximum bending corrugation somewhere in the middle of those two values. For t2G on hBN, we have checked that adding a fixed aligned hNB substrate below the existing hBN layer does not modify the qualitative behaviors and the system still shows local energy dips at the commensurate angles. We thus expect our conclusions based on free-standing systems to hold under more realistic experimental conditions when substrates are present.

\textbf{Conclusions} -- We have shown the tendency of double moire systems to spontaneously form commensurate moire patterns with rational $p/q$ moire length ratios and align their angles, as illustrated in the alternating twist trilayer graphene for $p=q=1$ and twisted bilayer graphene on hexagonal boron nitride for $p=1$, $q=1,2$ systems, and found that the effect quickly diminishes for $q > 1$.
The binding energy gained during the alignment of the twist angle near $\theta_{32} \simeq 1^{\circ}$, $1.5^{\circ}$ and $2^{\circ}$ is of the order of $\sim  0.2$, $0.17$, $0.13$~meV/atom respectively which is comparable in  magnitude with the energy differences resulting from the relative sliding of the top and bottom layers for commensurate moire geometries. These are several orders of magnitude larger than the sliding-dependent energy changes for incommensurate moire geometries indicating the existence of superlubricity as soon as the system is marginally twisted away from commensuration already for angles as small as $\sim 0.03^\circ$.
The rotation torque constants presented have been evaluated for different fixed bottom layers twist angles $\theta_{12}$. In experiments, once $\theta_{12}$ is fixed we propose that targeting $\theta_{32}$ angles that are equal or slightly larger than the value that yields commensurate double moire patterns will more easily tend to lock the system into moire commensuration,
while targeting a smaller angle may result in the system rotating back to the trivial zero-alignment commensurate phase upon thermal annealing. 
The qualitative conclusions based on the t3G and t2G/BN systems explored in this work are expected to apply for a variety of other twisted layered materials that will be investigated in future work.

\begin{acknowledgments}
\textbf{Acknowledgments} -- This work was supported by the Korean NRF through the Grants 
No. 2020R1A2C3009142 (N.L.),  2020R1A5A1016518 (J.A.), 2021R1A6A3A13045898 (Y.P.),
and Samsung Science and Technology Foundation Grant No. SSTF-BA1802-06 (J.J.).
We acknowledge computational support from KISTI Grant No. KSC-2022-CRE-0514 and by the resources of Urban Big data and AI Institute (UBAI) at UOS. J.J. also acknowledges support by the Korean Ministry of Land, Infrastructure and Transport(MOLIT) from the Innovative Talent Education Program for Smart Cities.
\end{acknowledgments}

\bibliography{all}

\section{Appendix A}
Commensurate moire superlattices can be built in general based on 4 integer indices using the conventions in Ref.~\cite{Hermann:2012dy}, as exemplified in Ref.~\cite{PhysRevB.106.115410} for twisted bilayer graphene systems.
For double moire systems, the same approach requires the definition of 6 integers, namely $(i,j,i^\prime,j^\prime,i^{\prime\prime},j^{\prime\prime})$. These integers define the following three transformation matrices where $\textbf{M}_i$ with $i=1,2,3$ correspond to L$_1$, L$_2$ and L$_3$ respectively:
\begin{eqnarray}
\begin{aligned}
& {\bm M}_1 =
\begin{pmatrix}
      i      &     j        \\
    - j      &     i + j
\end{pmatrix}, \\
& {\bm M}_2 =
\begin{pmatrix}
     i^\prime      & j^\prime       \\
    -j^\prime     & i^\prime + j^\prime
\end{pmatrix}, \\
&{\bm M}_3 =
\begin{pmatrix}
     i^{\prime\prime}     & j^{\prime\prime}       \\
    -j^{\prime\prime}     & i^{\prime\prime} + j^{\prime\prime}
\end{pmatrix}. \\
\end{aligned}
\label{eq:indice1}
\end{eqnarray}
These matrices relate the lattice vectors $\textbf{r}_1$ and $\textbf{r}_2$ to the lattice vectors $\textbf{a}_1$ and $\textbf{a}_2$ of the respective layers through
\begin{equation}
    \begin{pmatrix}
    \textbf{r}_1 \\
    \textbf{r}_2
    \end{pmatrix}
    = \textbf{M}_1\cdot
    \begin{pmatrix}
    \textbf{a}_1 \\
    \textbf{a}_2
    \end{pmatrix}
    = \textbf{M}_2\cdot
    \begin{pmatrix}
    \textbf{a}_1^{\prime} \\
    \textbf{a}_2^{\prime}
    \end{pmatrix}
    = \textbf{M}_3\cdot
    \begin{pmatrix}
    \textbf{a}_1^{\prime\prime} \\
    \textbf{a}_2^{\prime\prime}
    \end{pmatrix}
    \label{HermannEq}
\end{equation}
The lattice mismatch $\alpha_\text{mn}$ and twist angle $\theta_\text{mn}$ between the layers m and n can be related to these integers as follows
\begin{equation}
    \begin{aligned}
        \alpha_{12}&=\frac{|\textbf{a}_1|}{|\textbf{a}_1^{\prime}|} = \sqrt{\frac{{i^{\prime}}^2+{j^{\prime}}^2+i^{\prime} j^{\prime}}{i^2+j^2+i j}} , \\
        \alpha_{32}&=\frac{|\textbf{a}^{\prime\prime}_1|}{|\textbf{a}_1^{\prime}|} = \sqrt{\frac{{i^{\prime}}^2+{j^{\prime}}^2+i^{\prime} j^{\prime}}{{i^{\prime\prime}}^2+{j^{\prime\prime}}^2+{i^{\prime\prime}} {j^{\prime\prime}}}} ,\\
        \theta_{12} &= \theta_{1}-\theta_{2} = \cos^{-1}\left[\frac{ 2 i i^{\prime} + 2 j j^{\prime} + i j^{\prime} + j i^{\prime} }{2 \alpha_{12}(i^2+j^2+i j)}\right],\\
        \theta_{32} &= \theta_{3}-\theta_{2} = \cos^{-1}\left[\frac{ 2 i^{\prime\prime} i^{\prime} + 2 j^{\prime\prime} j^{\prime} + i^{\prime\prime} j^{\prime} + j^{\prime\prime} i^{\prime} }{2 \alpha_{32}({i^{\prime\prime}}^2+{j^{\prime\prime}}^2+{i^{\prime\prime}}{j^{\prime\prime}})}\right],\\
    \end{aligned}
     \label{eq:lattice}
\end{equation}
where we assume the middle L$_2$ layer as the untwisted reference layer.
In Table.~\ref{HermannIntegers}, we summarize the six integers used to generate the systems represented in Fig.~\ref{fig:t3G}(a) ~and~(b) for the t3G and the two t2G/BN systems. 

\begin{table*}[!htb]
    \begin{minipage}{.5\linewidth}
      \centering
 \resizebox{1.0\columnwidth}{!}{%
\begin{tabular}{|cccccc|}
\multicolumn{5}{c}{t3G} \\
\hline
$\theta_{12}$ ($^{\circ}$) &$\theta_{32}$ ($^{\circ}$) & ($i$ $j$ $i'$ $j'$ $i''$ $j''$) & $a_{L_3}$($\AA$) & \# atoms & $\lambda$ \\\hline
 1.084549 & 0.098591 & 341  330  330 341  331 340  & 2.460226   &   2026246 & $11 \lambda_1$  \\  
       & 0.299180 & 899  870  870 899  878 891  & 2.460278   &   14083050 & $29 \lambda_1$   \\  
       & 0.500560 & 403  390  390 403  396 397  & 2.460300   &   2829990  & $13 \lambda_1$  \\  
       & 0.699713 & 961  930  930 961  950 941  & 2.460291   &   16092466 & $31 \lambda_1$   \\  
       & 0.898632 & 1085  1050  1050 1085  1079 1056  & 2.460253   &   20513502 & $35 \lambda_1$   \\  
       & 1.049565 & 961  930  930 961  960 931  & 2.460204   &   16092846 & $31 \lambda_1$   \\  
       & \textbf{1.084549} & \textbf{31  30  30 31  31 30}  & \textbf{2.460190}   &   \textbf{16746} & $\mathbf{\lambda_1}$   \\  
       & 1.150275 & 1023  990  990 1023  1025 988  & 2.460162   &   18236534 & $33 \lambda_1$   \\  
       & 1.200741 & 868  840  840 868  871 837  & 2.460138   &   13129050 & $28 \lambda_1$   \\  
       & 1.491197 & 248  240  240 248  251 237  & 2.459963   &   1071810  & $8 \lambda_1$  \\  
       & 1.951946 & 155  150  150 155  159 146  & 2.459556   &   418722 & $5 \lambda_1$   \\ \hline 
 1.538500 & 0.099248 & 682  651  651 682  653 680  & 2.460244   &   7997326 & $31 \lambda_2$   \\  
       & 0.299135 & 792  756  756 792  763 785  & 2.460329   &   10784906 & $36 \lambda_2$   \\  
       & 0.498959 & 814  777  777 814  789 802  & 2.460384   &   11392218 & $37 \lambda_2$   \\  
       & 0.699314 & 242  231  231 242  236 237  & 2.460410   &   1006902  & $11 \lambda_2$  \\  
       & 0.901887 & 638  609  609 638  626 621  & 2.460405   &   6998394 & $ 29 \lambda_2$    \\  
       & 1.098945 & 154  147  147 154  152 149  & 2.460371   &   407758 & $7 \lambda_2$   \\  
       & 1.201971 & 704  672  672 704  697 679  & 2.460342   &   8521378  & $ 32\lambda_2$  \\  
       & 1.301824 & 286  273  273 286  284 275  & 2.460305   &   1406374 & $13 \lambda_2$   \\  
       & 1.398649 & 242  231  231 242  241 232  & 2.460263   &   1006942  & $11 \lambda_2$  \\  
       & 1.499055 & 858  819  819 858  857 820  & 2.460212   &   12657686 & $39 \lambda_2$   \\  
       & \textbf{1.538500} & \textbf{22  21  21 22  22 21}  & \textbf{2.460190}  &   \textbf{8322} & $\mathbf{\lambda_2}$   \\  
       & 1.577943 & 858  819  819 858  859 818  & 2.460167   &   12657842 & $39 \lambda_2$   \\  
       & 1.600032 & 550  525  525 550  551 524  & 2.460153   &   5201302  & $25 \lambda_2$  \\  
       & 1.700421 & 418  399  399 418  420 397  & 2.460087   &   3004326  & $19 \lambda_2$  \\  
       & 1.794869 & 132  126  126 132  133 125  & 2.460018   &   299606  & $6 \lambda_2$  \\  
       & 1.900421 & 374  357  357 374  378 353  & 2.459932   &   2405226  & $17 \lambda_2$  \\  
       & 1.999934 & 220  210  210 220  223 207  & 2.459844   &   832278  & $10 \lambda_2$  \\ \hline 
 2.004628 & 0.200433 & 170  160  160 170  161 169  & 2.460326   &   490182 & $10 \lambda_3$   \\  
       & 0.801831 & 255  240  240 255  246 249  & 2.460551   &   1102842 & $15 \lambda_3$   \\  
       & 1.202796 & 255  240  240 255  249 246  & 2.460551   &   1102842 & $15 \lambda_3$   \\  
       & 1.397199 & 561  528  528 561  551 538  & 2.460508   &   5337818  & $33 \lambda_3$  \\  
       & 1.603742 & 85  80  80 85  84 81  & 2.460431   &   122542 & $5 \lambda_3$    \\  
       & 1.700933 & 561  528  528 561  556 533  & 2.460384   &   5337998  & $33 \lambda_3$  \\  
       & 1.799055 & 663  624  624 663  659 628  & 2.460329   &   7455662  & $39 \lambda_3$  \\  
       & 1.899139 & 323  304  304 323  322 305  & 2.460265   &   1769586  & $19 \lambda_3$  \\  
       & 1.950459 & 629  592  592 629  628 593  & 2.460230   &   6710766  & $37 \lambda_3$  \\  
       & \textbf{2.004628} & \textbf{17 16 16 17 17 16}  & \textbf{2.460190}   &   \textbf{4902} & $\mathbf{\lambda_3}$   \\  
       & 2.056017 & 663  624  624 663  664 623  & 2.460150   &   7456022 & $39 \lambda_3$   \\  
       & 2.100064 & 357  336  336 357  358 335  & 2.460115   &   2161826  & $21 \lambda_3$  \\  
       & 2.198572 & 527  496  496 527  530 493  & 2.460030   &   4711026  & $31 \lambda_3$  \\  
       & 2.301520 & 459  432  432 459  463 428  & 2.459934   &   3573806 & $27 \lambda_3$   \\ \hline 
\end{tabular}
}
    \end{minipage}%
    \begin{minipage}{.5\linewidth}
      \centering
 \resizebox{1.0\columnwidth}{!}{%
\begin{tabular}{|cccccc|}
\multicolumn{5}{c}{t2G/BN Type I} \\
\hline
$\theta_{12}$ ($^{\circ}$) &$\theta_{32}$ ($^{\circ}$) &($i$ $j$ $i'$ $j'$ $i''$ $j''$) & $a_{L_3}$($\AA$) & \# atoms & $\lambda$ \\\hline
 0.560656 & 0.200227 & 812  812  812 840  817 835  & 2.460259   &   12143930 & $28 \lambda_4$  \\  
       & 0.400464 & 406  406  406 420  411 415  & 2.460298   &   3035950  & $ 14 \lambda_4$  \\  
       & 0.501637 & 1102  1102  1102 1140  1119 1123  & 2.460306   &   22366846  & $ 38 \lambda_4$  \\  
       & 0.546279 & 1131  1131  1131 1170  1150 1151  & 2.460308   &   23559530 & $ 39 \lambda_4$   \\  
       &\textbf{0.560656} & \textbf{58 58 58 60 59 59}  & \textbf{2.460308}   &   \textbf{61958}  & $\mathbf{2 \lambda_4}$  \\  
       & 0.583082 & 725  725  725 750  738 737  & 2.460308   &   9680938 & $ 25 \lambda_4$   \\  
       & 0.600704 & 812  812  812 840  827 825  & 2.460307   &   12143770  & $ 28 \lambda_4$  \\  
       & 0.700824 & 232  232  232 240  237 235  & 2.460300   &   991330  & $ 8 \lambda_4$  \\  
       & 0.800943 & 203  203  203 210  208 205  & 2.460286   &   758990 & $ 7 \lambda_4$   \\  
       & 0.897056 & 145  145  145 150  149 146  & 2.460265   &   387242  & $ 5 \lambda_4$  \\  
       & 1.000094 & 1073  1073  1073 1110  1106 1077  & 2.460235   &   21205546 & $ 37 \lambda_4$   \\  
       & 1.079783 & 783  783  783 810  809 784  & 2.460207   &   11292158  & $ 27 \lambda_4$  \\  
       & \textbf{1.121311} & \textbf{29 29 29 30 30 29}  & \textbf{2.460190}   &   \textbf{15490}   & $\mathbf{\lambda_4}$ \\  
       & 1.159974 & 841  841  841 870  871 840  & 2.460173   &   13027150 & $ 29 \lambda_4$   \\  
       & 1.229816 & 899  899  899 930  933 896  & 2.460140   &   14886094  & $ 31 \lambda_4$  \\  
       & 1.300703 & 725  725  725 750  754 721  & 2.460103   &   9681482  & $ 25 \lambda_4$  \\  
       & 1.495028 & 261  261  261 270  273 258  & 2.459981   &   1254762 & $ 9 \lambda_4$   \\  
       & 1.761923 & 203  203  203 210  214 199  & 2.459767   &   759098  & $ 7 \lambda_4$  \\  
       & 2.018092 & 145  145  145 150  154 141  & 2.459512   &   387322  & $ 5 \lambda_4$  \\ \hline 
\end{tabular}
}
\newline
\vspace*{0.77 cm}
\newline
\resizebox{1.0\columnwidth}{!}{%
\begin{tabular}{|cccccc|}
\multicolumn{5}{c}{t2G/BN Type II} \\
\hline
$\theta_{12}$ ($^{\circ}$) &$\theta_{32}$ ($^{\circ}$) & ($i$ $j$ $i'$ $j'$ $i''$ $j''$) & $a_{L_3}$($\AA$) & \# atoms & $\lambda$ \\\hline
 $-$0.579874 & 0.200227 & 840  784  840 812  835 817  & 2.460259   &   12145498  & $ 28 \lambda_5$  \\  
       & 0.537294 & 1440  1344  1440 1392  1417 1415  & 2.460308   &   35692418 & $ 48 \lambda_5$   \\  
       & \textbf{0.560656} & \textbf{60 56 60 58 59 59}  & \textbf{2.460308}   &   \textbf{61966}  & $ \mathbf{2 \lambda_5}$  \\  
       & 0.584017 & 1440  1344  1440 1392  1415 1417  & 2.460308   &   35692418 & $ 48 \lambda_5$   \\  
       & 0.600704 & 840  784  840 812  825 827  & 2.460307   &   12145338 & $ 28 \lambda_5$   \\  
       & 0.700824 & 240  224  240 232  235 237  & 2.460300   &   991458  & $ 8 \lambda_5$  \\  
       & 0.800943 & 210  196  210 203  205 208  & 2.460286   &   759088  & $ 7 \lambda_5$  \\  
       & 0.897056 & 150  140  150 145  146 149  & 2.460265   &   387292  & $ 5 \lambda_5$  \\  
       & 1.000094 & 1110  1036  1110 1073  1077 1106  & 2.460235   &   21208284  & $ 37 \lambda_5$  \\  
       & 1.079783 & 810  756  810 783  784 809  & 2.460207   &   11293616 & $ 27 \lambda_5$    \\  
       & \textbf{1.121311} & \textbf{30 28 30 29 29 30}  & \textbf{2.460190}   &   \textbf{15492} & $\mathbf{\lambda_5}$   \\  
       & 1.159974 & 870  812  870 841  840 871  & 2.460173   &   13028832 & $  29 \lambda_5$   \\  
       & 1.229816 & 930  868  930 899  896 933  & 2.460140   &   14888016 & $ 31 \lambda_5$   \\  
       & 1.300703 & 750  700  750 725  721 754  & 2.460103   &   9682732  & $ 25 \lambda_5$  \\  
       & 1.495028 & 270  252  270 261  258 273  & 2.459981   &   1254924  & $ 9 \lambda_5$  \\  
       & 1.761923 & 210  196  210 203  199 214  & 2.459767   &   759196   & $ 7 \lambda_5$ \\  
       & 2.018092 & 150  140  150 145  141 154  & 2.459512   &   387372  & $ 5 \lambda_5$  \\  
       & 2.099900 & 1530  1428  1530 1479  1434 1574  & 2.460238   &   40294166  & $ 51 \lambda_5$  \\  
       & 2.299955 & 1470  1372  1470 1421  1369 1521  & 2.460025   &   37197972  & $ 49 \lambda_5$  \\ \hline     
\end{tabular}
}
    \end{minipage} 
    \caption{Details about the commensurate cells that are used for our simulations on the t3G (left) and t2G/BN (right) systems where the first column contains $\theta_{12}$, the second column summarizes the $\theta_{32}$ for each of the corresponding $\theta_{12}$ values, the third column contains the six integers as defined in Ref.~\cite{PhysRevB.106.115410, Hermann:2012dy} where the first two integers control the lattice vectors of the first layer, the next two integers define the lattice of the second layer and the final two integers orient the lattice vectors of the top layer following Eq.~(\ref{HermannEq}). The fourth column contains the slightly strained lattice constant $a_{L_3}$ for the third layer which is different from the unstrained lattice constants of $2.4602$ \AA\ for $L_1$ and $L_2$. The fifth column contains the number of atoms and the final column represents the super-moire length or commensuration cell length $\lambda$ as a multiple of the commensurate cell moire length $\lambda_{i}$ when $\theta_{12} = \theta_{32}$, where $\lambda_1=129.97$~\AA, $\lambda_2=91.62$~\AA\ and $\lambda_3=70.32$~\AA\ for $\theta_{12}=1.0845^\circ$, $1.5385^\circ$ and $2.0046^\circ$ respectively for t3G and $\lambda_4=\lambda_5=125.71$~\AA\ for t2G/BN Type I and t2G/BN Type II. We highlight the commensurate angle configurations from the main text in bold while the other entries are commensurate approximations of the incommensurate angle combinations.}
    \label{HermannIntegers}
\end{table*}

\section{Appendix B}
\begin{figure}[tbhp]
\begin{center}
\includegraphics[width=1.0\columnwidth]{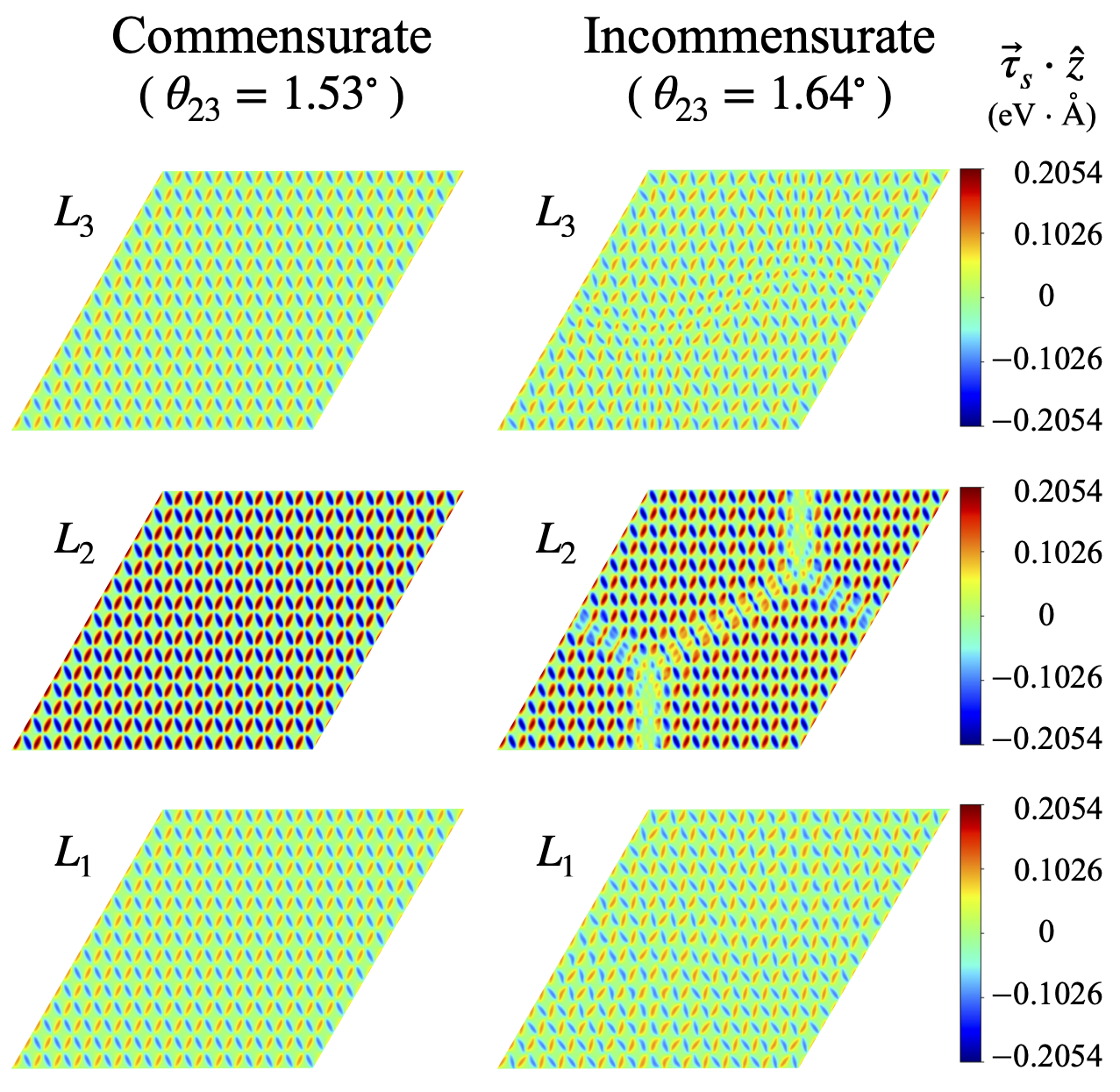}
\caption{
Local torque $\tau_s$ defined in Eq.~\ref{torqueMap} for each layer in t3G at the site $i\in$ sublattice $s$ are illustrated when the two moire interfaces are commensurate ($\theta_{12}=\theta_{32}=1.53^{\circ}$) (left) and incommensurate ($\theta_{12}=1.53^{\circ},\ \theta_{32}=1.64^{\circ}$) (right).  
We observe the presence of a super-moire pattern with a longer period for the incommensurate moire case that gives rise to the total energy differences with respect to the doubly commensurate moire case.
}
\label{torqueComparison}
\end{center}
\end{figure}
We illutrate here the difference between the local torque maps for a commensurate and an incommensurate moire system calculated through
\begin{equation}
    \boldsymbol{\tau}_s = (\textbf{r}_s-\textbf{r}_{cm}) \times \textbf{F}_s
    \label{torqueMap}
\end{equation}
where $s$ is the sublattice index and $cm$ refers to the center of mass of the dimer formed by the neighboring $A$ and $B$ sublattices,
and where $\textbf{F}_s$ is the interface component of the force acting on an atom extracted at the end of the LAMMPS minimization by subtracting the intralayer forces. 
The left panels show the local torque maps for commensurate moire pattern cases where the moire cell has been repeated 15 times for a more direct comparison with the right hand panels that have the same size. On the right panel, we illustrate the same torques for an incommensurate moire configuration modeled through a multiple moirons system with longer super-moire period. In the middle layer L$_2$ we see that a large region of the atoms feel the same torque as is seen for the commensurate phase, suggesting that within the region confined by long period triangular patches we largely recover the commensurate moire phase behavior seen for $\theta_{23} = 1.53^\circ$.
\end{document}